\documentclass[prd,aps,twocolumn,showpacs]{revtex4}

\usepackage{epsfig}

\newcommand{\be}{\begin{equation}}
\newcommand{\ee}{\end{equation}}
\newcommand{\bea}{\begin{eqnarray}}
\newcommand{\beas}{\begin{eqnarray*}}
\newcommand{\eea}{\end{eqnarray}}
\newcommand{\eeas}{\end{eqnarray*}}
\newcommand{\ba}{\begin{array}}
\newcommand{\ea}{\end{array}}

\newcommand{\lra}{\leftrightarrow}
\def\ls{\mathrel{\lower4pt\vbox{\lineskip=0pt\baselineskip=0pt
           \hbox{$<$}\hbox{$\sim$}}}}
\def\gs{\mathrel{\lower4pt\vbox{\lineskip=0pt\baselineskip=0pt
           \hbox{$>$}\hbox{$\sim$}}}}

\begin{document}


\title{Softly broken $\mu\leftrightarrow\tau$ symmetry in 
the minimal see-saw model}
\author{Juan Carlos G\'omez-Izquierdo\footnote{e-mail:jcarlos@fis.cinvestav.mx}
and Abdel P\'erez-Lorenzana\footnote{e-mail:aplorenz@fis.cinvestav.mx} }

\affiliation{Departamento de F\'{\i}sica, Centro
de Investigaci\'on y de Estudios Avanzados del I.P.N.\\
Apdo. Post. 14-740, 07000, M\'exico, D.F., M\'exico}

\date{September, 2007}

\begin{abstract}
Neutrino oscillations data indicates that neutrino mixings are consistent with
an apparent  $\nu_\mu - \nu_\tau$ exchange symmetry in neutrino mass  matrix. 
We observe that in the mininimally extended standard model with  the see-saw
mechanism,  one can impose  $\mu\leftrightarrow\tau$  symmetry at the tree
level  on all Lagrangian terms, but for the mass difference among $\mu$ and
$\tau$  leptons. In the absence of any new extra physics, this mass difference
becomes the only source for the breaking of such a symmetry, which induces, via
quantum corrections, small but predictable values for $\theta_{13}$, 
and for the  deviation of $\theta_{ATM}$ from maximallity. 
In the CP conserving case, 
the predictions only depend on  neutrino mass hierarchy and may
provide a unique way to test for new physics with neutrino experiments.
\end{abstract}

\maketitle

\section{Introduction}

Convincing evidence that neutrinos have mass and oscillate has been
provided along recent years by  neutrino oscillation experiments~\cite{maltoni}. 
In the standard framework,  only three weak 
neutrino species,  $\nu_e$; $\nu_\mu$ and $\nu_\tau$, are needed 
to consistently  describe the  experimental
results, with the  addition  of neutrino masses and mixings as  new
parameters to the standard model. 
Central idea  is
that neutrino mass eigenstates,
$\nu_{1,2,3}$,  and weak eigenstates are different, 
but they can be written as linear combinations of
each other by using a complex unitary matrix, $U$, as
$\nu_{\ell } = \sum_i U_{\ell i}\nu_{i}$, for $\ell=e,\mu,\tau$ and
$i=1,2,3$, where we refer only to left handed states.
A common parameterization for Majorana neutrinos 
of the $U$ matrix is given in terms of three angles
and three CP phases, such that  $U = V K$,  where  
$K={\rm diag}\{ 1,e^{i\phi_1},e^{i\phi_2}\}$, with $\phi_1$, $\phi_2$ the
physical CP-odd Majorana phases, and  the elements of the
$V$ mixing matrix  parameterized as~\cite{pmns} 
 \[
 V = \left(\ba{ccc} 
 c_{12} c_{13} & s_{12}c_{13} &  z^* \\
 -s_{12}c_{23} - c_{12}s_{23}z &
 c_{12}c_{23} - s_{12}s_{23}z & s_{23}c_{13} \\
 s_{12}s_{23} - c_{12}c_{23}z &
 -c_{12}s_{23} - s_{12}c_{23}z & c_{23}c_{13} \ea\right)~;
 \label{pmns}
 \]
where $c_{ij}$ and $s_{ij}$ stand for $\cos\theta_{ij}$ and $\sin\theta_{ij}$
respectively and $z=s_{13}e^{i\varphi}$. 
The kinematical  scales for the oscillation, on the other hand, 
are  given by the two mass squared
differences: the solar/KamLAND scale $\Delta m^2_{\odot}= \Delta m_{12}^2$;
and the atmospheric scale 
$\Delta m^2_{ATM}= |\Delta m_{23}|^2\approx|\Delta m_{13}|^2$. 
Combined analysis of all data~\cite{maltoni} indicates that at two sigma level
$\Delta m^2_{\odot} = 7.6~^{+0.5}_{-0.3} \times 10^{-5}~{\rm eV^2}$; 
$\Delta m^2_{ATM} = 2.4\pm 0.3 \times 10^{-3}~{\rm eV^2}$,
 whereas
$\sin^2\theta_{12}= 0.32~^{+0.05}_{-0.04}$, 
$\sin^2\theta_{ATM}=\sin^2\theta_{23} = 0.5~^{+0.13}_{-0.12}$,  
and  $\sin^2\theta_{13} \leq 0.033$.
Thus,  data is consistent with $\theta_{13}\approx0$, 
and $\theta_{ATM}\approx\pi/4$, which makes the Dirac CP phase, 
$\varphi$, hard to be measured. 
Current and new experiments on neutrino physics will explore how small and how
maximal, respectively, these mixing   are~\cite{nextexp}, down 
to the level of few times $10^{-2}$. 

Since the standard model 
was built on the assumption of zero neutrino masses, 
a fundamental question  at this point is whether neutrino
mass imply the existence of new physics, and what
such physics would be.  The answer, however, is not yet conclusive. 
It is possible to minimally extend the model by only
adding three singlet right handed neutrinos, $N_i$,  to
implement the
see-saw mechanism~\cite{see-saw,valle}, 
and accommodate data, without relaying in any new extra
ingredient.  This  makes, however, 
 the identification of any new extra physics 
from low energy phenomenology a difficult task. 
Above picture explains
very well the smallness of neutrino masses, but  
provides no understanding for the mixings. To provide such understanding, 
one usually is led to invoke theoretical arguments, and many ideas exist
nowadays in the literature. 

It has  already been observed that, in the limit with a null
$\theta_{13}$ and a maximal $\theta_{ATM}$, 
and on the basis where charge lepton
masses are diagonal,  the reconstructed neutrino see-saw  mass matrix,
$M_{\ell\ell'}=\sum_{i=1}^3U_{\ell i}^* m_{i}U_{\ell' i}$,  posses a
$\nu_\mu - \nu_\tau$ exchange symmetry~\cite{mt}. This has inspired a large
number of theoretical studies~\cite{numodels}.   
Remarkably, imposing the suggested  
$\mu \leftrightarrow\tau$ symmetry  is very well  possible within the minimal
see-saw extension of the standard model, and it is our goal  
to show that 
the simplest realization of
this idea provides a  perfectly falsifyable
model, with specific predictions that can easily 
be proved wrong by future neutrino
data. 
Our findings, however, would show that with 
these minimal ingredients the prediction for both $\theta_{13}$, and the
deviation of $\theta_{ATM}$ from maximallity are rather much smaller than 
the forthcoming
experimental sensitivities. Nevertheless, there is positive outcome, 
our results stablish a comparative point of reference 
such as to take any possible 
measurement of a  non zero value for those mixing parameters  
in near future experiments  as clear
indications for the existence of new physics.  

It is not difficult to see that $\mu \leftrightarrow \tau$ is already a  
flavor symmetry in the standard model, 
but for the charged lepton mass terms, where
clearly $m_\tau\neq m_\mu$.  Thus, we propose to  treat  
$\mu\leftrightarrow \tau$ as a softly broken symmetry of the minimal see-saw
model. 
Therefore, at tree level,  all
physics not directly related to $m_{\mu,\tau}$ would be described by 
the symmetric limit, allowing us to fix the free parameters of the model
at low energy.  
Nevertheless, quantum corrections shall communicate
the symmetry breaking to the neutrino sector~\cite{weinberg}. 
In particular, one loop corrections will
already produce small deviations to $\theta_{ATM}$ from $\pi/4$, and a non
zero $\theta_{13}$. 
Because the model has not extra unknown ingredients, 
one can make definite predictions for these physical
observables in terms of symmetric level results. Those are the main points
we want to discuss in what follows.  

\section{The minimal  $\mu\leftrightarrow\tau$ model}

The  model we will explore considers, first,  the minimal see-saw
extension that includes  three right handed neutrinos, with  
all additional
Lagrangian terms that are consistent with the standard model symmetries,
\be 
h_{\ell}\bar L_\ell H\ell_R + 
y_{\ell \ell'}\bar L_\ell \tilde H N_{\ell'} +  (h.c.) +
(M_R)_{\ell\ell'}\bar N^c_\ell N_{\ell'}~,
\label{ssm}
\ee
where sum over indices should be understood. Here, $L_\ell$ stands for the
standard lepton doublets and $H$ for the Higgs field. In order to implement
$\mu\leftrightarrow \tau$ symmetry in a meaningful way, we have chosen to work
in the basis where the charge lepton Yukawa couplings, and so their masses, 
are diagonal and real. Also, we have chosen right handed neutrinos to carry 
lepton number, and properly identified the index. 
It is worth mentioning that if
$N_{i\neq\ell}$  were not subjected to 
$\mu\leftrightarrow \tau$ symmetry, as defined below, then, 
neutrino Yukawa couplings would became
such that $y_{\mu\, i} = y_{\tau\, i}$, under $\mu\leftrightarrow \tau$
symmetry,
regardless of the chosen basis
for $N_i$. Following this implied degeneracy of second and
third rows on the Dirac mass matrix, the left handed massless neutrino state 
$\nu'= (\nu_\mu -\nu_\tau)/\sqrt{2}$ arises. Clearly, this 
corresponds to the third mass eigenstate in an inverted hierarchy
scenario (similar results were recently found in Ref.~\cite{baba}). 
 
Next, to realize the symmetry, we  require both 
Yukawa couplings and Majorana
mass matrix to be 
invariant under $\mu\leftrightarrow \tau$ exchange: 
$L_\mu\leftrightarrow L_\tau$;  $\mu_R\leftrightarrow \tau_R$; 
and $N_\mu\leftrightarrow N_\tau$.
One can then proceed with
the diagonalization of the Mass matrices. 
However, the analysis for the low energy phenomenology is
simplified by first implementing the see-saw mechanism, and observing that 
$\mu\leftrightarrow \tau$ symmetry  also holds  for 
the effective left handed neutrino mass matrix, 
$M= - m_DM_R^{-1}m_D^T$, with $m_D$ the Dirac neutrino mass matrix.
Here, the symmetry expresses itself by two
simple conditions on matrix elements: $M_{e\mu} = M_{e\tau}$  and
$M_{\mu\mu}=M_{\tau\tau}$.
Thus, the most general tree level form for $M$  should be
 \be
 M=M_{\mu\leftrightarrow\tau}=\left(\ba{ccc} 
 m_{ee}^0   & m_{e\mu}^0    & m_{e\mu}^0   \\[.5ex]
 m_{e\mu}^0 & m_{\mu\mu}^0  & m_{\mu\tau}^0\\[.5ex]
 m_{e\mu}^0 & m_{\mu\tau}^0 & m_{\mu\mu}^0
 \ea\right)~.
\label{msym}
 \ee
Diagonalization of such a mass matrix is rather simple. 
We find the mass eigenvalues:
\bea
m_1 &=&  m_{ee}^0 - \sqrt{2}\,\tan\theta_{12}\,m_{e\mu}^0~; \nonumber\\
m_2 &=&  m_{ee}^0 + \sqrt{2}\,\cot\theta_{12}\,m_{e\mu}^0~; \nonumber\\
m_3 &=& m_{\mu\mu}^0 - m_{\mu\tau}^0~,
\label{mis}
\eea 
and get for the mixing angles 
$\theta_{ATM}= \pi/4$, and $\theta_{13}=0$, whereas 
the solar mixing angle is given by
\be 
\tan2\theta_{12} = 
\frac{\sqrt{8}\,m_{e\mu}^0}{m_{\mu\mu}^0+ m_{\mu\tau}^0 -m_{ee}^0}~,
\label{tansol}
\ee
Since
$\sin\theta_{13}=0$,  
the Dirac CP phase, $\varphi$, gets  undefined. 
Thus, in this model only the 
two CP Majorana phases may exist at the symmetric limit. 
To keep our present discussion
simple,  we will assume them to be zero along the analysis,
and so we shall take all mass parameters in Eq.~(\ref{msym}) to be real.
The analysis including CP phases will be presented elsewhere. 

It is worth noticing  that the
see-saw mass matrix in Eq.~(\ref{msym})
is  described by only four parameters,  
which can be 
entirely fixed by the following four 
low energy observables: the solar mixing
($\theta_{12}$), the mass hierarchy ($m_3$),  solar scale
$\Delta m^2_{\odot}=m_2^2-m_1^2$, 
and the atmospheric scale that we can take as 
$\Delta m^2_{ATM}=\frac{1}{2}|\Delta m_{13}^2 +\Delta m_{23}^2|$.
Of course, extra parameters yet exist for the whole theory 
[see Eq.~(\ref{ssm})]. They belong to the high energy right handed neutrino 
sector and cannot be  fixed from these results.
However, as we will show below, those parameters will not be  
required to make further predictions for the low energy physics.

\section{Soft breaking of $\mu\leftrightarrow\tau$ symmetry}

Exact $\mu\leftrightarrow \tau$ symmetry would also imply 
that $h_\mu=h_\tau$, 
which gives the wrong result $m_\mu=m_\tau$. 
This is the only  place where the 
symmetry 
is being  explicitely broken. Henceforth, we will take  
$h_\mu\neq h_\tau$, a choice that of course respects all gauge symmetries,
and by definition it is expected to be valid at any energy. 
We shall not assume
any dynamical origin for such a difference on the Yukawa couplings
in order to keep the model  truly minimal.
Notice that, as a matter of fact, 
all leptonic kinetic terms in the standard model, 
$i\bar L_\ell \gamma^\mu D_{\mu L} L_\ell + 
i\bar\ell_R\gamma^\mu D_{\mu R}\ell_R$
with $D_{L,R}^\mu$ the corresponding  covariant derivatives, 
are  invariant under 
$\mu\leftrightarrow \tau$ exchange due to the 
universality of gauge interactions.
In contrast, out of the Lagrangian terms given in Eq.~(\ref{ssm}), the 
charged lepton Yukawa couplings now have the  form
\be h_e \bar{L}_e H e_R + 
h_\mu\bar{L}_\tau H \mu_R + (h_\mu+\delta h)\bar{L}_\tau H\tau_R + h.c.~, 
\ee
whereas all other terms  
remain symmetric. 
Therefore, 
the whole tree level Lagrangian
of the model can be written as 
${\cal L} = {\cal L}_{\mu\leftrightarrow \tau} + 
(\delta h  \bar L_\tau H\tau_R + h.c.)$.
Here, ${\cal L}_{\mu\leftrightarrow \tau}$ contains all symmetric terms,
whereas the last term will generate the mass term $\delta m\, \bar\tau\tau$,
upon standard gauge symmetry breaking.  
The last can be seen as a soft term whose role is to correct the mass 
of the tau lepton.

The breaking down of 
$\mu\leftrightarrow \tau$ symmetry will be communicated to all other sectors of
the model via weak interactions~\cite{weinberg}. Particularly,
after including one-loop 
quantum corrections, muon-tau mass difference will generate a
splitting in the symmetry conditions of the see-saw mass matrix, such that 
one should rather have $M_{e\tau}\neq M_{e\mu}$
and $M_{\tau\tau}\neq M_{\mu\mu}$, where the departure would  be expected
to be small due to the $W$ mass supressions, but calculable (see Ref.~\cite{doi}
for a general analysis of quantum corrections on $M_\nu$). 
Notice that our calculation will be done at the low scale where observable 
neutrino mass parameters are being measured. No running from renormalization
group equations is included, which is also known to produce very mild effects
on the mixings that concern us here (for related works 
considering renormalization group corrections 
on neutrino mixings see for instance references in~\cite{rad1,rad2}).
As a matter of fact, at one loop order  it is easy
to see that the only diagram contributing to neutrino mass corrections 
which is not  $\mu\leftrightarrow \tau$ invariant
is the one given in Fig. 1, which explicitely involves the exchange of charge
leptons through $W$ couplings. The violation of the symmetry conditions would 
imply both, a non zero value for
$\theta_{13}$ and the departure of $\theta_{ATM}$ from maximal, which as we
will show next are completely predictable and correlated.

\begin{figure}[ht]
\epsfxsize=150pt
\epsfbox{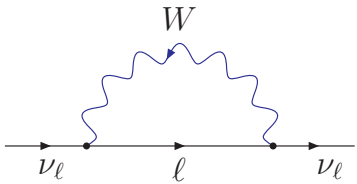}
\caption{1-loop diagram that communicates the breaking 
of $\mu\lra\tau$ symmetry to neutrino mass matrix}
\end{figure}

Therefore, after including one loop corrections the neutrino mass matrix 
gets the more general form
\be
 M=\left(\ba{ccc} 
 m_{ee}   & m_{e\mu}    & m_{e\tau}   \\
 m_{e\mu} & m_{\mu\mu}  & m_{\mu\tau}\\
 m_{e\tau} & m_{\mu\tau} & m_{\tau\tau}
 \ea\right)~,
\label{msoft}
\ee
which  is now written in terms of 
the corrected mass parameters given by 
$m_{\ell\ell'} = m_{\ell\ell'}^{0} + 
I_{\ell\ell''}m_{\ell''\ell'}^{0} + m_{\ell\ell''}^{0}I_{\ell''\ell'}$, with
$I_{\ell\ell'}$ the one loop finite contributions to mass terms that come 
from all possible one loop diagrams. 
The former matrix can be written as
$M=M_{\mu\leftrightarrow\tau} + \delta M$ 
where the symmetric part, $M_{\mu\leftrightarrow\tau}$ 
has a similar parameterization as in Eq.~(\ref{msym}), although
now in terms of corrected masses. 
On the other hand 
$\delta M$ encodes the only two  
symmetry breaking conditions,  that at the lower order are 
respectively given by
$\delta M_{e\tau} \equiv m_{e\tau} - m_{e\mu} \approx m_{e\mu}^0\Delta I$, 
and 
$\delta M_{\tau\tau} \equiv m_{\tau\tau} - 
m_{\mu\mu} \approx 2\,m_{\mu\mu}^0\Delta I$,
where  $\Delta I \equiv I_{\tau} - I_{\mu}$  with $I_{\ell}$ the one loop
contributions obtained from the diagram in Fig. 1 for the corresponding
charged lepton $\ell$.
A quite lengthy calculation shows that 
\be 
\Delta I \approx \frac{3g^2_W}{32 \pi^2}
\left[\left(\frac{m_\tau}{M_W}\right)^2\ln\left(\frac{m_\tau}{M_W}\right)-
(\tau\rightarrow\mu)\right]~,
\ee
which gives $\Delta I\approx -7.68\times 10^{-6}$.

Due to the smallness of $\Delta I$, 
the neutrino mass matrix in Eq.~(\ref{msoft}) 
can be diagonalized considering expressions 
up to linear order
corrections in $\Delta I$. 
Interestingly enough, the  effect on neutrino masses and solar
mixing enters as  
a slight modification of previous formulas that consists on 
the sole replacing of  $m_{e\mu}$ and $m_{\mu\mu}$
by the average values  
$\overline m_{e\mu}=\frac{1}{2}(m_{e\mu} +m_{e\tau})$, and 
${\overline m_{\mu\mu}}=\frac{1}{2}(m_{\mu\mu} +m_{\tau\tau})$,
respectively, in Eqs.(\ref{mis}) and (\ref{tansol}), such that
we now get
\bea
m_1 &\approx&  m_{ee} - \sqrt{2}\,\tan\theta_{12}\,{\overline m_{e\mu}}~; 
\nonumber\\
m_2 &\approx&  m_{ee} + \sqrt{2}\,\cot\theta_{12}\,{\overline m_{e\mu}}~; 
\nonumber\\
m_3 &\approx& {\overline m_{\mu\mu}} - m_{\mu\tau}~,\nonumber\\
\tan2\theta_{12} &\approx& 
\frac{\sqrt{8}\,{\overline m_{e\mu}}}
{{\overline m_{\mu\mu}}+ m_{\mu\tau} -m_{ee}}~.
\label{mis2}
\eea
As already observed, one can invert these equations  to express the
involved neutrino mass parameters  in terms of  
neutrino observables and  $m_3$ as the hierarchy parameter, by using 
$|m_1|\approx \sqrt{m_3^2\mp\Delta m_{ATM}^2 -\frac{1}{2}\Delta m_{\odot}^2}$, 
and 
$|m_2|\approx \sqrt{m_3^2\mp\Delta m_{ATM}^2 +\frac{1}{2}\Delta m_{\odot}^2}$, 
where  the minus (plus) sign corresponds to normal (inverted) hierarchy.
After some algebra one gets
\bea 
m_{ee}&\approx& m_1 \cos^2\theta_{12} + m_2\sin^2\theta_{12}~; \nonumber\\
{\overline m_{e\mu}} &\approx& \frac{1}{\sqrt{8}}\sin2\theta_{12}\, (m_2-m_1)~;
\nonumber\\
{\overline m_{\mu\mu}} &\approx& 
\frac{1}{2}(m_1 \sin^2\theta_{12} + m_2\cos^2\theta_{12} +m_3)~;\nonumber\\
m_{\mu\tau} &\approx& 
\frac{1}{2}(m_1 \sin^2\theta_{12} + m_2\cos^2\theta_{12} -m_3)~.
\label{msol}
\eea

Next, we get the following predictions for 
other  mixings, at the lower order, 
\be
\sin\theta_{13} \approx -\frac{\Delta I}{\sqrt{2}}
\left(\frac{m_3\,{\overline m_{e\mu}}}
{{\overline m_{e\mu}}^2+m_{\mu\tau}\,m_*}\right) 
\label{sth13}
\ee
with $m_*= m_3 - m_{ee}$, 
and 
\be 
\sin\alpha = \frac{\Delta I }{2}
\left(\frac{{\overline m_{e\mu}}^2 + {\overline m_{\mu\mu}}\, m_*}
{{\overline m_{e\mu}}^2 + m_{\mu\tau}\,m_* }\right)  
\label{salpha}
\ee
for the deviation of $\theta_{ATM}$ from maximality, where we have defined 
$\alpha= \theta_{ATM}-\pi/4$.
In those formulas we have conveniently  approximated  
$m_{e\mu}^0\approx \overline m_{e\mu}$, and 
$m_{\mu\mu}^0 \approx \overline m_{\mu\mu}$, 
by using the tree level formulas in
Eqs.~(\ref{mis}) and (\ref{tansol}), 
and comparing them  with Eqs.~(\ref{mis2}).  
Thus, by committing an small error of order $\sim (\Delta I)^2$, 
this approach allows us to express $\sin\theta_{13}$ and $\sin\alpha$ 
in terms of all the tree level quantities  obtained from neutrino 
low energy observables  given in Eq.~(\ref{msol}). 
Predicted values, however,
depend not only on the hierarchy but also on the relative signs among the mass
eigenvalues. This becomes clear if we study, for instance, 
the expressions (\ref{sth13}) and 
(\ref{salpha}) in the limit of almost degenerate neutrinos, for which the
relative sign among $m_1$ and $m_2$ may enhance or suppress the 
contribution of $\overline m_{e\mu}$. 
These are our findings: 

First, we get the approximated formulas
\bea
\sin\theta_{13} &\approx&
A\cdot\frac{m_3^2\,\sin{2\theta_{12}}}{\Delta m^2_{ATM}}\,\Delta I~,\nonumber\\
\sin\alpha&\approx& 
\mp B\cdot \frac{2\, m_3^2}{\Delta m^2_{ATM}}\,\Delta I~,
\label{apform}
\eea
where $A$ and $B$ coefficients are given as
\begin{itemize}
\item[(i)]
 $A= \Delta m^2_\odot/\Delta m^2_{ATM}$, and $B=1$ for all $m_{1,2,3}>0$;

\item[(ii)] $A= \pm 1$ and $B=c_{12}^2$ for $m_1<0$ and $m_{2,3}>0$; and

\item[(iii)] $A= \mp 1$ and $B=s_{12}^2$ for $m_{1,3}>0$ but $m_2<0$;
\end{itemize}
where, the upper (lower) signs corresponds to normal (inverted) hierarchy.
Notice that second and third cases predict 
$|\sin\theta_{13}|\approx 5\times 10^{-4}(m_3/0.4~eV)^2$, which is larger than
case (i) prediction  by a factor of 
about 33, whereas
in all cases $|\sin\alpha|\approx B\cdot 10^{-3} (m_3/0.4~eV)^2$, and so, 
it  comes about the same order. 

Finally, (iv), for $m_3>0$ and $m_{1,2}<0$ one obtains
\bea
\sin\theta_{13} &\approx&
-\frac{\Delta m_\odot^2\,\sin{2\theta_{12}}}{16\, m_3^2}\,\Delta I ~,\nonumber\\
\sin\alpha&\approx& 
\mp\frac{\Delta m_{ATM}^2}{8\, m_3^2} \Delta I ~,
\eea
which indicate that smaller values respect to other cases would be expected.
Notice that the inverse squared $m_3$ mass dependence is only valid in the 
almost degenerate limit we are considering. 
For the hierarchical case former formulas become
$\sin\theta_{13}\approx\sin\alpha \sim - \Delta I/2 $ 
for normal hierarchy, whereas 
$\sin\alpha\approx\Delta I/2$ and 
$\sin\theta_{13}\sim m\, \Delta m_\odot^2/(\Delta m_{ATM}^2)^{3/2}$ for
inverted hierarchy. 

From above results, it is clear that 
$\theta_{ATM}$ should be on the first (second) octant for normal (inverted)
hierarchy, whereas experimental determination of the 
sign of $\theta_{13}$ would 
discriminate  cases (i) and (ii) from other ones. A measurement of
$|\sin\alpha|$ would finally  resolve the scenario. 
To get the whole picture of the parameter region that experiments should reach
to test the present model, we present in Fig. 2 the two sigma regions 
for our predicted values of $|\sin\theta_{13}|$ and $|\sin\alpha|$
for cases (i) to (iii). Results from case (iv) are simply out of range. 
Plot points were obtained from a direct numerical calculation using
Eqs.~(\ref{sth13}) and (\ref{salpha}) for $m_3<0.4$~eV. 
Absolute values are used to depict all results together. Notice that 
hierarchy makes a
clear difference  for low $m_3$ values in case (i) 
 for which the upper band on the lower left corresponds to normal
hierarchy. Differences also exist on other cases
for small $m_3$, although they are less evident.

\begin{figure}[ht]
\centerline{
\epsfxsize=180pt
\epsfbox{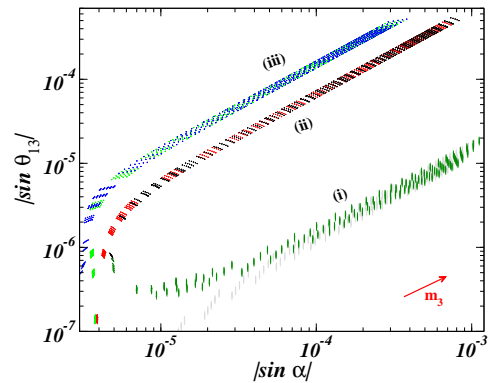}
}
\caption{
Two sigma regions for 
$|\sin\theta_{13}|$ and  $|\sin\alpha|$ predicted from the soft
$\mu\leftrightarrow \tau$  model, for both  normal 
and inverted hierarchies. Results correspond to cases (i) to (iii) as discussed
in the text, for $m_3<0.4$~eV running from small to larges values,
as indicated.}
\end{figure}

From the figure, we notice that predicted values are rather small,  as we
expected. Both $\sin\theta_{13}$ and $\sin\alpha$ are below $10^{-3}$, which is
clearly far below the expected  sensitivity of the near future forthcoming
experiments. Thus we would have to wait for a distant future experiment
to test the  depicted parameter zone to get 
a positive signal for the model. 
Nevertheless, if experiments determine values
out of these regions, which could happen in the near future, 
that would be a clear indication that, either,
(a) new physics beyond the standard model is involved in the breaking of  
$\mu\leftrightarrow\tau$, and the generation of the $\delta M$ corrections, or
(b) $\mu\leftrightarrow\tau$ is not a good symmetry to guide  model
building. The symmetry seems so natural that, from our point of view, 
it would be more likely that the first option would be the correct one
in such a case.

\section{Lepton Flavor violation processes}

Before closing our discussion it is worth mentioning 
another direct implication of our model for lepton physics. 
Since the breaking effects of $\mu\leftrightarrow\tau$ symmetry are 
rather small, lepton number violation processes would be
ruled in a good approximation by this symmetry. Besides, 
due to the lack of beyond standard model physics in our model, 
only $W$ exchanged diagrams would contribute to such
processes, and because they are proportional to neutrino squared masses and
mixings, they are predicted to be extremely small. 
This provides another clear
way to determine the existence of new physics if any observable effect
associated to lepton flavor violating processes is detected in near future
experiments.
    
In particular, the decay ratios for $\mu\rightarrow e\gamma$ and
$\tau\rightarrow e\gamma$  at one loop order would be~\cite{eliezer}  
\be 
\Gamma(\ell\rightarrow e\gamma)
\approx 
\frac{\alpha}{4\pi^4} G_F^2 \sin^22\theta_\odot (\Delta m_\odot^2)^2 m_\ell~,
\ee
for $\ell=\mu,\tau$ and $\alpha=e^2/4\pi$. Therefore one gets the relation
$\Gamma(\mu\rightarrow e\gamma)/\Gamma(\tau\rightarrow e\gamma)
\approx m_\mu/m_\tau\approx 0.06$, which means 
$B(\mu\rightarrow e\gamma)/B(\tau\rightarrow e\gamma)\approx
m_\mu\Gamma_\mu/m_\tau\Gamma_\tau \sim 8\cdot 10^{-8}$ 
for the corresponding branching
ratios.

Notice that the overall factor 
$G_F^2 (\Delta m_\odot^2)^2 $ is already too small to provide any visible
effect within the reach of current and near future experimental sensitivities.  
Indeed,  a
straightforward calculation for the branching ratio, say for instance for
$\mu\rightarrow e\gamma$, gives
\be 
B(\mu\rightarrow e\gamma)
\approx 
\frac{48 \alpha}{\pi}\sin^22\theta_\odot \frac{(\Delta m_\odot^2)^2}{m_\mu^4}~,
\ee
which is about $5\cdot 10^{-41}$. 
Tau decay into muon-gamma, on the other hand,  has the rate 
\be 
\Gamma(\tau\rightarrow \mu\gamma)
\approx \frac{(\Delta m_{ATM}^2)^2}{\sin^22\theta_\odot (\Delta m_\odot^2)^2}
\cdot \Gamma(\tau\rightarrow e\gamma)~.
\ee
Thus, the branching ratio for $\tau\rightarrow\mu\gamma$, although  enhanced 
by a factor of thousand respect to that for $\tau\rightarrow e\gamma$, yet  
remains far from reachable too.

In comparison,  $\mu\rightarrow eee$ and $\tau\rightarrow eee$
decays are expected to be yet more suppressed~\cite{eliezer}. Simple
$\gamma\rightarrow ee$ insertion on previous processes will amount to an extra
suppression factor of order $\alpha$ over above results, without altering 
the relation amoung the decay rates. Thus one would also get 
 $\Gamma(\mu\rightarrow eee)/\Gamma(\tau\rightarrow eee)
\approx m_\mu/m_\tau$.

\section{Concluding remarks}

Summarizing, we have presented the  minimal see-saw model that realizes 
$\mu\leftrightarrow\tau$ symmetry at tree level in all Lagrangian terms,  but
for the muon  and tau mass difference, which in the absence of any extra new
physics, becomes the only breaking source for the symmetry. The model  predicts,
through quantum corrections, small values for $\theta_{13}$ and for the
deviation of $\theta_{ATM}$ from maximal, which, on the absence of CP
violation, only  depend on neutrino mass
hierarchy. We also notice that lepton flavor violation processes are 
controlled by the $\mu\leftrightarrow\tau$ symmetry. However, the main
contributions to such precesses come out to be suppressed by a factor of
$(G_F \Delta m_{\odot}^2)^2$, which make them too small to be reachable by any
near future experiment. We stress that even though above results are difficult to
be tested in any forthcoming experiment, they may have a positive outcome: 
since we
are working in the minimal  model that extends the standard model to
include neutrino physics parameters, our numerical findings provide a
clean  point of comparison with the experiment, such as to claim that  
any positive experimental signal for,
either, a non zero $\theta_{13}$ or $\alpha$ mixing, 
or for any of the described lepton flavor violation processes, 
would be a clear indication for the existence
of new physics, beyond the
present setup.


\acknowledgments
This work was supported in part by CONACyT, M\'exico, grant number 54576.


\end{document}